\documentclass[slac_one]{revtex4}
\usepackage{graphicx}
\usepackage{fancyhdr}
\usepackage{lscape}
\def\beq{\begin{equation}}
\def\eeq#1{\label{#1}\end{equation}}
\def\eeqn{\end{equation}}

\def\leqn#1{(\ref{#1})}

\let\bar=\overbar

\def\VEV#1{\left\langle{ #1} \right\rangle}

\def\lsim{\mathrel{\raise.3ex\hbox{$<$\kern-.75em\lower1ex\hbox{$\sim$}}}}
\def\gsim{\mathrel{\raise.3ex\hbox{$>$\kern-.75em\lower1ex\hbox{$\sim$}}}}

\def\L{{\cal L}}

\def\half{\frac{1}{2}}

\def\del{\partial}
\def\Dslash{\not{\hbox{\kern-4pt $D$}}}
\def\dslash{\not{\hbox{\kern-2pt $\del$}}}

\def\Pl{{\mbox{\scriptsize Pl}}}

\def\ee{e^+e^-}

\def\msb{{\bar{\ssstyle M \kern -1pt S}}}

\def\s#1{\widetilde{#1}}

\makeatletter
\def\section{\@startsection{section}{0}{\z@}{5.5ex plus .5ex minus
 1.5ex}{2.3ex plus .2ex}{\large\bf}}
\def\subsection{\@startsection{subsection}{1}{\z@}{3.5ex plus .5ex minus
 1.5ex}{1.3ex plus .2ex}{\normalsize\bf}}
\def\subsubsection{\@startsection{subsubsection}{2}{\z@}{-3.5ex plus
-1ex minus  -.2ex}{2.3ex plus .2ex}{\normalsize\sl}}

\renewcommand{\@makecaption}[2]{%
   \vskip 10pt
   \setbox\@tempboxa\hbox{\small #1: #2}
   \ifdim \wd\@tempboxa >\hsize     
       \small #1: #2\par          
     \else                        
       \hbox to\hsize{\hfil\box\@tempboxa\hfil}
   \fi}

 \def\citenum#1{{\def\@cite##1##2{##1}\cite{#1}}}
 
\newcount\@tempcntc
\def\@citex[#1]#2{\if@filesw\immediate\write\@auxout{\string\citation{#2}}\fi
  \@tempcnta\z@\@tempcntb\m@ne\def\@citea{}\@cite{\@for\@citeb:=#2\do
    {\@ifundefined
       {b@\@citeb}{\@citeo\@tempcntb\m@ne\@citea\def\@citea{,}{\bf ?}\@warning
       {Citation `\@citeb' on page \thepage \space undefined}}%
    {\setbox\z@\hbox{\global\@tempcntc0\csname b@\@citeb\endcsname\relax}%
     \ifnum\@tempcntc=\z@ \@citeo\@tempcntb\m@ne
       \@citea\def\@citea{,}\hbox{\csname b@\@citeb\endcsname}%
     \else
      \advance\@tempcntb\@ne
      \ifnum\@tempcntb=\@tempcntc
      \else\advance\@tempcntb\m@ne\@citeo
      \@tempcnta\@tempcntc\@tempcntb\@tempcntc\fi\fi}}\@citeo}{#1}}
\def\@citeo{\ifnum\@tempcnta>\@tempcntb\else\@citea\def\@citea{,}%
  \ifnum\@tempcnta=\@tempcntb\the\@tempcnta\else
  {\advance\@tempcnta\@ne\ifnum\@tempcnta=\@tempcntb \else\def\@citea{--}\fi
    \advance\@tempcnta\m@ne\the\@tempcnta\@citea\the\@tempcntb}\fi\fi}
\makeatother

\def\Title#1{\begin{center} {\Large #1 } \end{center}}
\def\Author#1{\begin{center}{ \sc #1} \end{center}}
\def\Address#1{\begin{center}{ \it #1} \end{center}}
\def\andauth{\begin{center}{and} \end{center}}
\newcommand\pubblock{\rightline{\begin{tabular}{l} SLAC--PUB--11479\\
        LBNL--58837\\    September, 2005 \end{tabular}}}
\newenvironment{Abstract}{\begin{quotation} \begin{center}
                       ABSTRACT
     \end{center}\bigskip  }{\end{quotation}}
\newenvironment{Presented}{\begin{quotation} \begin{center} 
             PRESENTED AT\end{center}\bigskip 
      \begin{center}\begin{large}}{\end{large}\end{center} \end{quotation}}
\def\SLAC{Stanford Linear Accelerator Center\\
    Stanford University, Stanford, California 94309 USA}
\def\doeack{\footnote{Work supported by the US Department of Energy,
                     contract DE--AC02--76SF00515.}}
\def\LBL{Department of Physics and Lawrence Berkeley Laboratory \\ 
  University of California, Berkeley, California 94720 USA}
\def\lblack{\footnote{Work supported by the US Department of Energy,
                     contract DE--AC02--05CH11231.}}

\begin{document}
\begin{titlepage}
\pubblock

\vfill
\Title{The Role of the ILC in the Study of Cosmic Dark Matter}
\vfill
\Author{Marco Battaglia\lblack}
\Address{\LBL}
\andauth
\Author{Michael E. Peskin\doeack}
\Address{\SLAC}
\vfill
\begin{Abstract}
Though there is strong evidence that dark matter is a major component of the
universe, most aspects of dark matter are completely mysterious.  We do not 
know what dark matter is, and we do not know how it is distributed in our
galaxy.  To  resolve these and related questions, we will need information
both from particle physics and from astrophysics.  In this article, we will 
describe a path toward the solution of the problems of dark matter, and we 
will highlight the important role that the ILC has to  play in this study.  
\end{Abstract}
\vfill
\begin{Presented}
2005 International Linear Collider Workshop\\
Stanford University, Stanford, California  \\   18-22 March, 2005
\end{Presented}
\vfill
\end{titlepage}
\def\thefootnote{\fnsymbol{footnote}}
\setcounter{footnote}{0}
\newpage
\mbox{\null}
\newpage

\title{{\small{2005 International Linear Collider Workshop - Stanford,
U.S.A.}}\\ 
\vspace{12pt}
The Role of the ILC in the Study 
       of Cosmic Dark Matter} 

%

\author{Marco Battaglia}
\affiliation{Dept. of Physics, University of California  and LBNL,
Berkeley, CA 94720 USA}
\author{Michael E.  Peskin}
\affiliation{SLAC, Stanford CA 94309 USA}

\begin{abstract}
Though there is strong evidence that dark matter is a major component of the
universe, most aspects of dark matter are completely mysterious.  We do not 
know what dark matter is, and we do not know how it is distributed in our
galaxy.  To  resolve these and related questions, we will need information
both from particle physics and from astrophysics.  In this article, we will 
describe a path toward the solution of the problems of dark matter, and we 
will highlight the important role that the ILC has to  play in this study.  
\end{abstract}

\maketitle

\thispagestyle{fancy}


\section{INTRODUCTION}

Dark matter is well established as a major component of the universe.
We see its gravitational influence at the scales of galaxies and clusters
of galaxies and in the dynamics of the plasma that emitted the photons now 
seen as the cosmic microwave background.  These measurements give a 
consistent estimate that dark matter makes up about 20\% of the total 
energy density of the universe.

Almost every other property of dark matter, however, is a mystery.  We do not
know what dark matter is made of. The various explanations of dark matter
in terms of elementary particles range from particles of mass 10$^{-5}$ eV 
(axions) through particles of mass 10$^{18}$ GeV (`WIMPzillas'), and even to
particles of earth or Jupiter mass (primordial black holes).  The paradigm
of cosmic structure formation by cold dark matter appears to agree with 
observations on very large scales, but it is controversial whether this 
model predicts too large concentrations of mass at the center of galaxies
and too many substructures and small companions for the Milky Way. 

To address these problems, we need to observe dark matter particles in the
galaxy,  and to understand those observations, we need to measure the 
properties of dark matter particles in high-energy physics experiments.
Without both halves of the story, we will not be able to reconstruct the 
full picture.

For many of the possibilities for the identity of the dark matter particle, 
we may never be able to assemble all of this information.  However, there
is a general class of  candidate dark matter
particles for which we can find out experimentally
both what they are and where they are.  It is possible that the study 
of these particles could take us over the complete path to the concrete
understanding of dark matter. In this talk, we will 
sketch the particle physics aspects of this study, and the central role
that the ILC will play in it.

\section{WHY THE WIMP MODEL DESERVES SPECIAL ATTENTION}

Among the many particle physics candidates for dark matter, one should
receive pride of place.  This is the WIMP, which we define to be a massive
neutral stable particle that was once in thermal equilibrium in the early
universe.  

The initial condition of thermal equilibrium allows us to compute
the present  cosmic
 density of such a particle, assuming knowledge of the particle's
interactions and the extrapolation of standard cosmology back to a temperature
comparable to the particle's mass. To perform this computation, one
integrates the Boltzmann equation to follow the density of WIMPs as the 
universe cools to temperatures much lower than the WIMP mass.  The WIMPs 
drop out of thermal equilibrium, and, because of the expansion of the 
universe, their density becomes so small that further annihilation has a 
negligible effect.  The resulting density is the `relic density' of the 
WIMP. To 10\% accuracy, this density is given 
by the relation~\cite{TurnerScherrer}
\beq
\Omega_{\chi}h^2 = {s_0\over \rho_c/h^2} \left( {45\over \pi g_*}\right)^{1/2}
          { x_f \over m_\Pl }{1\over \VEV{\sigma v}}
\eeq{OmegaN}
where $s_0$ is the current entropy density of the universe, $\rho_c$ is the
critical density, $h$ is the (scaled) Hubble constant, $g_*$ is the number of 
relativistic degrees of freedom at the time that the dark matter particle
goes out of thermal equilibrium, $m_\Pl$ is the Planck mass, $x_f \approx 25$,
and  $\VEV{\sigma v}$ is the thermal average of the 
dark matter pair annnihilation cross section times the relative velocity.
Most of these quantities are numbers with
large exponents.  However, combining them and equating the result to 
$\Omega_\chi \sim 0.2$, we obtain
\beq
        \VEV{\sigma v} \sim 1 \ \mbox{pb}
\eeq{findsigmav}
Interpreting this in terms of a mass, using $\VEV{\sigma v} = 
\pi \alpha^2/8m^2$, we find $m = 100$~GeV.

This is a remarkable result, because it places the WIMP at 
a mass scale where we already expect to find physics beyond the Standard 
Model.  It is highly suggestive that the new physics that is responsible
for the breaking of electroweak symmetry also gives rise to a WIMP that
is responsible for the dark matter.  In fact, it is more than suggestive.
In every model of electroweak symmetry breaking, it is possible
to add a discrete symmetry that makes the lightest new particle stable.
Often, this discrete symmetry is required for other reasons.  For example,
in supersymmetry, the conserved R parity is needed to eliminate rapid 
proton decay.  In other cases, such as models with TeV-scale extra 
dimensions, the discrete symmetry is a natural consequence of the underlying
geometry.  As long as it is generic that the lightest stable particle is
neutral, we have a WIMP that is guaranteed to give---to order of 
magnitude---the correct cosmic density to agree with observations.

If the WIMP model of dark matter is preferred by theory, it is also 
preferred by experiment, or, at least, by experimenters.
  Many experiments are now trying to observe 
dark matter from the galaxy, and more are proposed for the near future.
These include `direct detection' experiments, in which one observes
the dark matter as scattering events in sensitive underground 
detectors, and `indirect detection', which one observes the
products of dark matter annihilation.
With a few exceptions, such as the Livermore axion search 
experiment~\cite{axion}, all
of these experiments
require that the dark matter particle is a heavy neutral particle
with weak-interaction cross sections.

Thus, there are very likely to be WIMP candidates for dark matter.  Only
for these candidates can we perform the crucial experiments that identify
the dark matter and tell us the distribution of dark matter in the galaxy.
It would be wonderful if a single type of WIMP could account for all of the 
dark matter in the universe.  But whether this is true or not, we ought to 
settle the issue experimentally.  Let us discuss how this can be done.

\section{THE IMPORTANCE OF THE LHC IN DARK MATTER STUDIES}

There is another common feature of WIMP models based on models of electroweak
symmetry breaking.  Since these models must somehow generate the quark 
masses, there are typically new particles with nonzero color.  In 
supersymmetry, for example, we have the squarks.  These particles carry the
conserved discrete quantum number and so eventually decay to the WIMP.
These colored states can be produced copiously in proton-proton collisions.
Thus, with the extra assumption that such particles exist and have masses 
below about 2~TeV, the LHC will produce huge numbers of WIMPs.  

To a first approximation, the cross sections for production of pairs of 
colored particles at the LHC depend only on the mass of the particle.  For
colored particles of mass below 1 TeV, these cross sections are tens of
pb.  In models with a conserved discrete quantum number, the pair-production
events will contain several jets coming from the decays of the primary 
particles, plus two WIMPs that will exit a particle physics detector
unobserved. These events have the `jets plus missing energy' signature that
is often considered to be characteristic of supersymmetry.  In fact, this 
signature appears in all models of WIMP dark matter that contain 
new colored particles satisfying the assumptions just described.

\begin{figure*}[t]
\centering
\includegraphics[width=8.2cm]{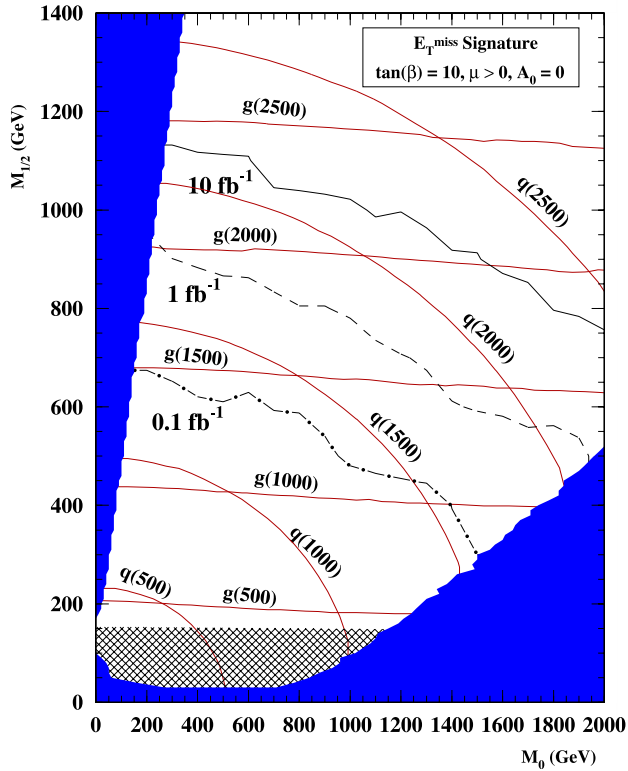}
\caption{Regions of the mSUGRA parameter space giving the discovery
of the missing-energy signature at the LHC with various levels of
 integrated luminosity, from~\cite{Tovey}.
  To use this result more generally, follow the marked curves of 
          constant squark and gluino masses.} \label{fig:LHCSUSY}
\end{figure*}

Since the event rates for missing energy events are determined mainly by the
primary particle masses, we can estimate these from the results for 
supersymmetry, which have been worked out in detail.  From
Fig.~\ref{fig:LHCSUSY}~\cite{Tovey}, we see 
that, for primary particle masses below
1~TeV, the integrated luminosity required to discover the missing energy
signature is amazingly low, about 100 pb$^{-1}$, or 1\% of the LHC first-year
design luminosity.  It is often noted that it may be a long time before 
we see signs of the Higgs boson at the LHC. But for WIMP dark matter, the 
situation is completely different.  We will know almost immediately whether
the LHC is producing a WIMP dark matter candidate with a mass at the 
weak interaction scale.

However, this statement comes with an important qualification.  Although 
it will be obvious that the LHC is producing a particle candidate for
dark matter,  there are scenarios where it might be very difficult to
determine, even qualitatively, the identity of that candidate. Consider, 
for example, the 
four possible models illustrated on the left-hand side of
Fig.~\ref{fig:ambiguous}.  This figure
shows the decay chain of the colored primary in models of  supersymmetry
in which the WIMP might be  a neutralino or a sneutrino and in models of
TeV-scale extra dimensions in which the WIMP might be the partner of the 
$U(1)$ gauge boson or of a neutrino~\cite{CMSCH}.   
In all four cases, the observable
leptons and jets in the decay chain are the same.  Because at least two 
unobservable WIMPs are produced in each event, it is not possible to 
reconstruct the detailed kinematics.  Within specific models,
characteristic features of the model can 
 be used at the LHC to exclude some of the
possibilities~\cite{BDDKM,SW}.  But it is likely that a number of options
will 
remain viable until  it is possible to do experiments of a much more 
incisive type.

Within specific models with a small number of parameters, the LHC data
can be used in a powerful way to estimate the dark matter relic 
density~\cite{Polesello}.  However, this method goes only so far if we
do not know the model.  In addition, as well will see in a moment, many
models of dark matter---in particular, models in much
 of the space of supersymmetry
preferred by the WMAP result---pose special problems for the LHC in 
carrying out model-indepedent predictions of the relic 
density~\cite{Battaglia:2004mp}.

\section{QUALITATIVE STUDIES OF DARK MATTER AT THE ILC}

The problem just described is one that the ILC is well suited to solve.
The ILC will provide  point-like collisions, tunable centre-of-mass 
energy, and the availability of powerful analysers such as beam polarisation. 
These tools will give us 
the ability to systematically determine the mass spectrum and 
the interactions of new particles. Even if the ILC may not be 
able to produce the 
heaviest new states, the WIMP annihilation 
cross section depends most strongly on 
the lightest particles in the new sector. 
As long as the ILC can reach the first
new particle thresholds, it will make the precision measurements 
of those particles that are 
most important for predicting the dark matter density.

In models of WIMP dark matter, there is typically a lightest visible particle
that decays to the WIMP.  In supersymmetry models, this is a slepton or a 
wino; in extra-dimensional models, it is the Kaluza-Klein recurrence of a 
lepton or a gauge boson.  The various options are distinguished if we can 
identify the spin and $SU(2)\times U(1)$ quantum numbers of this particle.
This plays to a strength of $\ee$ annihilation. The cross section for 
pair-production in $\ee$ annihilation through a virtual $\gamma$ and $Z$
has a characteristic energy- and 
angular-dependence for each value of the spin, and  the normalization of the 
cross section is chacteristic function of the electric charge and weak 
isospin.  This measurement then cleanly separates the various cases.  The 
angular distributions in the decay of the particles provide a check of the 
spin identification.  

\begin{figure*}[t]
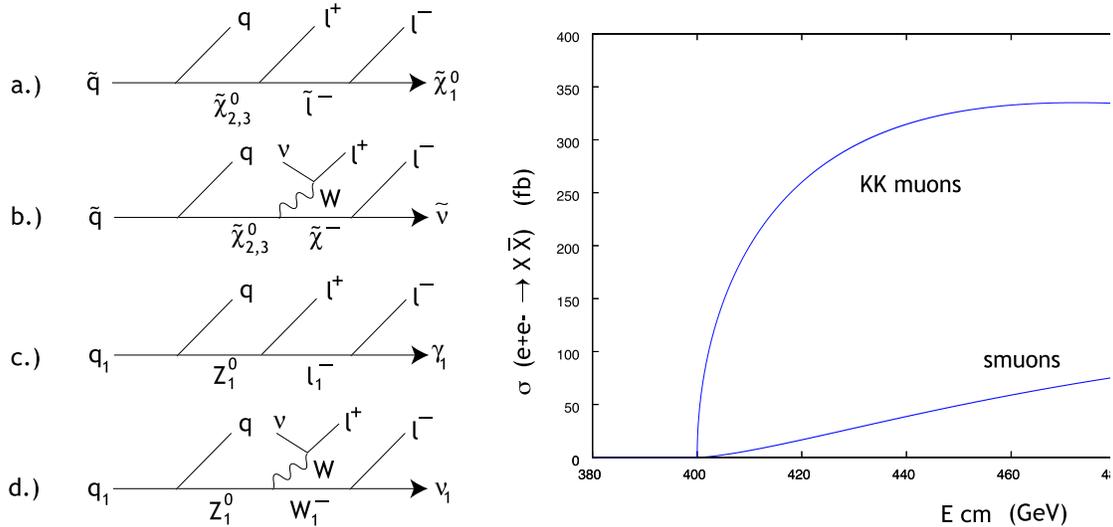

\centering
\includegraphics[width=6cm]{newWIMPoptions.eps}\qquad 
\includegraphics[width=8cm]{newDMcross.eps}\ 
\caption{Left: Four possible physics models for missing-energy events at the 
       LHC; Right, discrimination of the models (a) and (c) by a total cross 
section measurement at the ILC.} \label{fig:ambiguous}
\end{figure*}

To illustrate this, we show on the 
right-hand side of  Fig.~\ref{fig:ambiguous} 
the cross sections as 
a function of energy for the lightest particle that decays to a muon and 
a WIMP in the models shown in Fig~\ref{fig:ambiguous}(a) and (c).  Whereas
these models have essentially the same phenomenology at the LHC, we see 
that they are distinguished in a obvious way at the ILC. 

\section{QUANTITATIVE ANALYSIS OF WIMP DARK MATTER}

Once we have used
the ILC results to make the qualitative identification of the model, we 
are ready to move on to the next level of analysis. A heavy neutral particle
observed at accelerators is a candidate for the cosmic dark matter, but this
in no way proves that the dark matter is actually composed of this particle.
There are three types of observations, though, that would go a long way 
toward providing that proof.  First, we should observe the WIMP in an 
astrophysical experiment and check that the particle mass seen there is the 
same at that observed at accelerators.  Second, we should determine the 
parameters of the WIMP model well enough to provide a microscopic prediction
of the WIMP relic density.  This can be compared to the dark matter density
obtained, for example, from the cosmic microwave background.  Third, we should
check that the microscopic model gives a pattern of WIMP cross sections that
is consistent with the result of direct and indirect detection experiments.
This last point raises astrophysical questions that we will discuss further 
in Section 8. 

There are many reasonable scenarios in which these tests would fail.  The
WIMP  observed at the LHC and the ILC
could make up only a fraction of the cosmic dark matter.  The
WIMP could decay to a `super-WIMP'~\cite{SuperWIMPs} with very 
small astrophysical cross sections, leading to a decreased dark matter density 
and zero signal in direct and indirect detection experiments.  The WIMP 
could be produced in the early universe through a mechanism that operates out
of thermal equilibrium, leading to a larger density, or the density could be
diluted by entropy production after the dark matter density is established,
that is, in the period between $10^{-10}$ sec and 10 sec after the Big Bang.
In all of these cases, the results of the experiments that we will describe 
provide a starting point for analyzing the difficulty and charting out the
full theory of dark matter.  And, in case the two values of the DM density 
will actually agree, it would be striking evidence 
that we would have understood 
the origin of dark matter.  This would  be  a great triumph for 
both particle physics and cosmology.
Before guesssing that this hypothesis is too optimistic, 
one should 
remember that the simplest hypotheses for nuclear physics in 
the early universe 
beautifully explain the primordial element abundances~\cite{Steigman}.

Dark matter detection experiments might measure the mass of the dark matter
particle to 10-20\% accuracy.  The LHC should already measure the mass 
of the WIMP to comparable accuracy, setting up a first confrontation of 
particle physics and astrophysical results.  To go beyond this level, it is 
necessary to determine the WIMP interaction cross sections.
It is not so easy to experimentally 
determine the cross sections of an  unobservable particle.
  To see that this can be done, we would now like to specialize
to the case of supersymmetry and  neutralino WIMP dark matter, for which we 
have carried out 
explicit model analyses.

\section{FIXING THE NEUTRALINO RELIC DENSITY}

Typical models in the parameter space of minimal supersymmetry predict too 
large densities of neutralino dark matter.  In such models, the annihilation
cross sections are suppressed, either because the particles exchanged are 
heavy or because the important couplings for annihilation are suppressed by 
small mixing angles.  Models of supersymmetry that produce the observed
dark matter relic density do so because some specific mechanism leads 
enhanced neutralino annihilation.  The enhancement of this cross section
might be due, for example, to the presence of light sleptons, leading to 
$\chi\chi \to \ell^+\ell^-$, to sizable gaugino-Higgino mixing, enhancing
$\chi\chi \to W^+W^-$ and $\chi\chi \to ZZ$, 
or to an accidental degeneracy $m_A \approx 2 m_\chi$, 
leading to annihilation through the $A^0$ Higgs
boson as an $s$-channel resonance.

\begin{table*}[t]
\centering
\begin{tabular}{l|cccccc|c| r r} 
   Point &  $m_0$ & $m_\half$ & $\tan\beta$ & $A_0$ & $\mbox{sign}\ mu$ & $m_t$
       &    reference    &  \qquad  $\Omega_\chi h^2$ &  ILC accuracy \\ \hline
   LCC1 &   100 &  250 &   10  & $-100$ & $+$ & 178 & \cite{ILCLHC} & 
                      0.193 &             $\pm$  1.0\% \\
   LCC2 & 3280 & 300 & 10 & 0 &  $+$ & 175 &   \cite{Gray}  & 
                        0.110 &     $\pm$ 3.2\%  \\ 
   LCC3 &  210 & 360 &  40 &  0 & $+$ & 178 & \cite{Dutta} & 
                         0.057 &   $\pm$ 7.5\%   \\
   LCC4 &  380 & 420 &  53 & 0   & $+$ & 178 & \cite{BattagliaParis} &
                0.106 &   $\pm$ 4.9\% \\
\end{tabular}
\caption{mSUGRA parameter sets for four illustrative models of neutralino
     dark matter.  Masses are given in GeV.}
\label{tab:LCCpoints}
\end{table*}

Each of these mechanisms can be elucidated by specific experiments on the
supersymmetric particles.
  Essentially, we must determine with high accuracy the masses and 
couplings of the specific particles that enter the key annihilation reactions,
and we must exclude the importance of  competing annihilation processes.
This requires, for the first goal, precision measurements on the 
lightest particles in the 
supersymmetry spectrum and, for the second goal, the model-independent
exclusion of the possibility that other particles are comparably light.
Both goals are very difficult for the LHC.  The LHC can often make precise
measurements of some particles in the spectrum, but it is difficult for 
the LHC experiments to assemble the complete set of parameters needed to 
reconstruct annihilation cross section.  And, it is typical that supersymmetry
spectra contain light particles that are very difficult to observe in the
hadron collider environment.  The ILC, in contrast, provides just the right
setting to obtain both types of measurements.  Again, it is not necessary
for the ILC to match the energy of the LHC, only that it provides enough
energy to see the lightest charged particles of the new sector.

To illustrate these considerations, a number of specific points in the
parameter space of minimal supergravity (mSUGRA) have been chosen for 
detailed study. Points in mSUGRA are specified by four parameters and a 
sign.  The parameters chosen are listed in Table I.  The supersymmetry 
spectra associated with these parameters were computed using 
ISAJET 7.69~\cite{ISAJET}.
Predictions for the dark matter relic density were computed using
Micromegas~1.3~\cite{Micromegas}; DarkSUSY~\cite{DarkSUSY} gives similar
results.

The LCC points are chosen to illustrate the various scenarios for the 
neutralino relic density, in models in which the lightest charged 
supersymmetric particles can be studied at 500 GeV in the center of mass.
LCC1 is chosen as the point SPS1a whose collider observables are 
studied in great detail  in \cite{ILCLHC}. At this point, $t$-channel
exchange of light sleptons dominates the annihilation cross section.
 LCC2 is chosen as a point in 
the `focus point region' of mSUGRA, with very heavy squarks
and sleptons.  At this point, the annihilation  to $W^+W^-$ and 
$Z^0Z^0$ is the 
dominant mode. The ILC measurement capabilities at this point have been 
studied in \cite{Gray}.  LCC3 is a point with relatively heavy sleptons, but
with the lightest slepton ($\tilde\tau_1$) having a mass close to that of the
neutralino.  In this circumstance, the $\tilde\tau_1$ is almost as 
abundant as the neutralinos at the time that the relic abundance is 
established, and the dominant processes for supersymmetric particle 
annihilation are $\tilde\chi_1^0 \tilde\tau_1$ and $\tilde\tau_1 \tilde\tau_1$
annihlation.  This situation is call `coannihilation'~\cite{Seckel}.  
 The ILC measurement capabilities at this point have been 
studied in \cite{Dutta}, and the specific issue of measuring the 
$\tilde\tau_1$ mass with precision has been addressed in \cite{Richard}.
LCC4 is a point at which the $A^0$ Higgs boson is relatively close to the 
neutralino pair threshold, so that annihilation through the $A^0$ resonance
dominates the annihilation process.  The ILC measurement capabilities at 
this point have been studied in \cite{BattagliaParis}. 

The ILC analyses just cited are based on parametric simulation that 
includes realistic 
detector performances and effects of the ILC beam characteristics. 
The studies have assumed
that the ILC will be able to provide collisions at centre-of-mass energies 
from 0.3~TeV to 0.5~TeV with an integrated luminosity of 500~fb$^{-1}$ in the 
first phase of operation. Several of the studies have also assumed
 a second phase of running at 1 TeV in the center of mass, with an 
additional data set of 1~ab$^{-1}$. We show in Table~\ref{tab:constraints}
the estimated accuracies on masses and mass differences derived from these
studies.

\begin{table*}
\centering
\begin{tabular}{l|cccc}
Observable & \quad LCC1 \quad  & \quad LCC2 \quad & \quad LCC3 \quad 
& \quad LCC4 \quad \\ \hline
$M(\tilde \chi^0_1)$ & $\pm$ 0.05 & $\pm$ 0.7 & $\pm$ 0.1 & $\pm$ 1.4 \\
$M(\tilde e_R) $ &  $\pm$ 0.05 & - & $\pm$ 1.0 & $\pm$ 0.6 \\
$M(\tilde \tau_1)$ &   $\pm$ 0.3 & - & $\pm$ 0.5 & $\pm$ 0.9  \\
$M(\tilde \tau_2)$ &  $\pm$ 1.1 & - & - & - \\
$M(\tilde \chi^+_1) $   & $\pm$ 0.55  & $\pm$ 0.7 & & $\pm$ 0.6 \\
$M(\tilde \tau_1) - M(\tilde \chi^0_1)$ &  & - & $\pm$ 1.0 & $\pm$ 1.0 \\
$M(\tilde \tau_2) - M(\tilde \chi^0_1)$ &  &  & $\pm$ 1.1 & \\
$M(\tilde\chi^0_2)-M(\tilde\chi^0_1)$ & $\pm$ 0.7
 & $\pm$ 0.4 & $\pm$ 2.0 & $\pm$ 1.8\\
$M(\tilde\chi^0_3)-M(\tilde\chi^0_1)$ & & $\pm$ 0.3 & $\pm$ 0.5 & $\pm$ 2.0 \\
$M(\tilde \chi^+_1)-M(\tilde \chi^0_1)$ &  & $\pm$ 0.3 & & $\pm$ 2.0 \\
$M(\tilde \chi^+_2)-M(\tilde \chi^+_1)$ &  & & $\pm$ 2.0 & $\pm$ 2.0 \\
$M(A^0) $ &   &  &  & $\pm$ 0.8 \\
$\Gamma(A^0) $ &   &  &  & $\pm$ 1.2 \\ 
\end{tabular}
\caption{Summary of the main mass constraints from the ILC for the four 
benchmark points.}
\label{tab:constraints}
\end{table*}

To assess the abilities of the LHC and ILC collider experiments to predict
the dark matter relic density in these models, we have carried out
broad scans of the parameter space of supersymmetric models.  Previous 
studies (e.g., \cite{Polesello}) have converted collider measurements
to predictions for $\Omega_\chi h^2$ using the assumption that the underlying
model belongs to the 4-parameter space of mSUGRA models.  We believe that this
assumption is much too restrictive to realistically assess the impact of 
a set of collider measurements.  In our analysis, we have described
the benchmark points in terms of 24
 effective MSSM parameters at the electroweak scale. These parameters sweep
out the most general models within the MSSM in which 
flavor and CP are conserved.  We have
then carried out a scan over these 24 parameters
 to find the full set of MSSM models
 that would be consistent with measurements made
at each benchmark point.
We have used two strategies for the scans. 
In the first, we have made a flat scan in which the MSSM parameters
have been independently varied over wide ranges. Each scan point 
has been weighted by the likelihood that the masses $\{m_j\}$ and other
spectral information at the benchmark point can be reproduced by
the predictions of the scan point within the errors:
\beq
    \L(P_i) = \prod_j \exp \left\{ -{(m_j(P_i) - m_j(P_0))^2\over 
                    2 \sigma_j^2 } 
      \right\}\ ,
\eeq{likelihood}
where $P_i$ is the scan point, $P_0$ is the benchmark point, and 
$\sigma_j$ is the expected experimental accuracy of the measurement.
The ILC accuracies for measurements at the various benchmark points, 
taken from the studies referred to in the previous paragraph, are displayed 
in Table~\ref{tab:constraints}.
These weights can then be used
to build a probability density function for the dark matter relic density,
and for cross sections of interest in astrophysics.

 The flat scan method offers a even 
sampling of the 
parameter phase space, but it is quite inefficient, because the high
accuracies of measurements at the ILC select out models in narrow regions
of the parameter space.  A more efficient way to select points is 
the Markov Chain Monte Carlo algorithm~\cite{Berg, BaltzGondolo}.
 In collaboration with Baltz and Wizansky, we have 
adopted this strategy to compute probability densities for the prediction
of dark matter properties.  In this method, one 
steps from one point $P_i$ in the MSSM parameter space to 
the next point $P_{i+1}$ if the likelihood \leqn{likelihood} increases;
if the likelihood decreases, one makes the step with the probability
$\L(P_{i+1})/\L(P_i)$.  This rule produces an ensemble of points that are 
generated with probabilty proportional to $\L(P_i)$.
 The Markov Chain method offers a more effective sampling 
of the parameter phase space compared to the flat scan.
A detailed discussion of our analysis and its 
results will be given in~\cite{BBPW}.

\begin{figure*}[t]
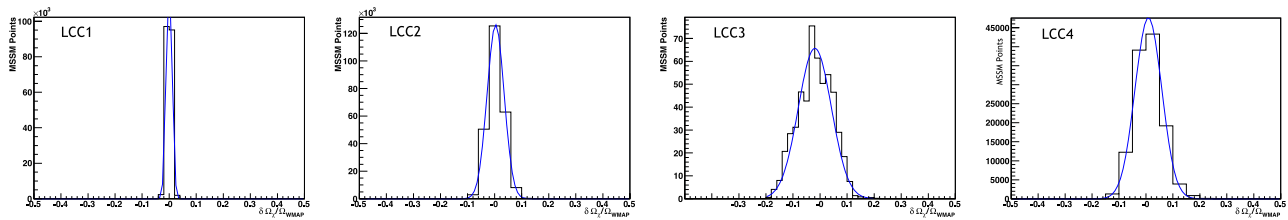

\centering
\includegraphics[width=4.0cm]{newplotLCC1.eps}\quad 
\includegraphics[width=4.0cm]{newplotLCC2.eps}\quad 
\includegraphics[width=4.0cm]{newplotLCC3.eps}\quad 
\includegraphics[width=4.0cm]{newplotLCC4.eps} 
\caption{Probabilility distribution of predictions for the neutralino
 relic density $\Omega_\chi h^2$ based on expected measurements at the 
LHC and at the ILC, in the four supersymmetry models presented in
 Table I.} \label{fig:scans}
\end{figure*}

This analysis leads to estimates of the precision of the prediction of the 
neutralino relic density from the ILC measurements for the four 
benchmark points.  These estimates are given in the right-hand column 
of Table~\ref{tab:LCCpoints}.  The accuracies range from   1\% in the 
most straightforward case to 7.5\% in the model with coannihilation.
The scan data for the four points, and  fits to Gaussian distributions, are
shown in Fig.~\ref{fig:scans}.

\section{COMPARISON OF ILC AND LHC}

Using the methods described in the previous section, it is possible to 
compare predictions for the neutralino relic density from measurements
at the ILC to the determinations from LHC measurements.  The two analyses
can be carried out in parallel, by writing the suite of supersymmetry
spectroscopy measurements expected at each benchmark point, constructing
the likelihood function, and then following one of the scan strategies
described above.

It is important to note that, to apply this analysis to the LHC, 
we must begin from the assumption
that  the underlying physics model is supersymmetry.  As we have 
emphasized in Section 3, this assumption would need to be justified by 
data from the ILC.   To keep this in mind, we have labeled the curves from 
the LHC analysis `LHC (after Q)', that is, after qualitative identification
of the model. 

\begin{figure*}[t]
\centering
\includegraphics[width=9.2cm]{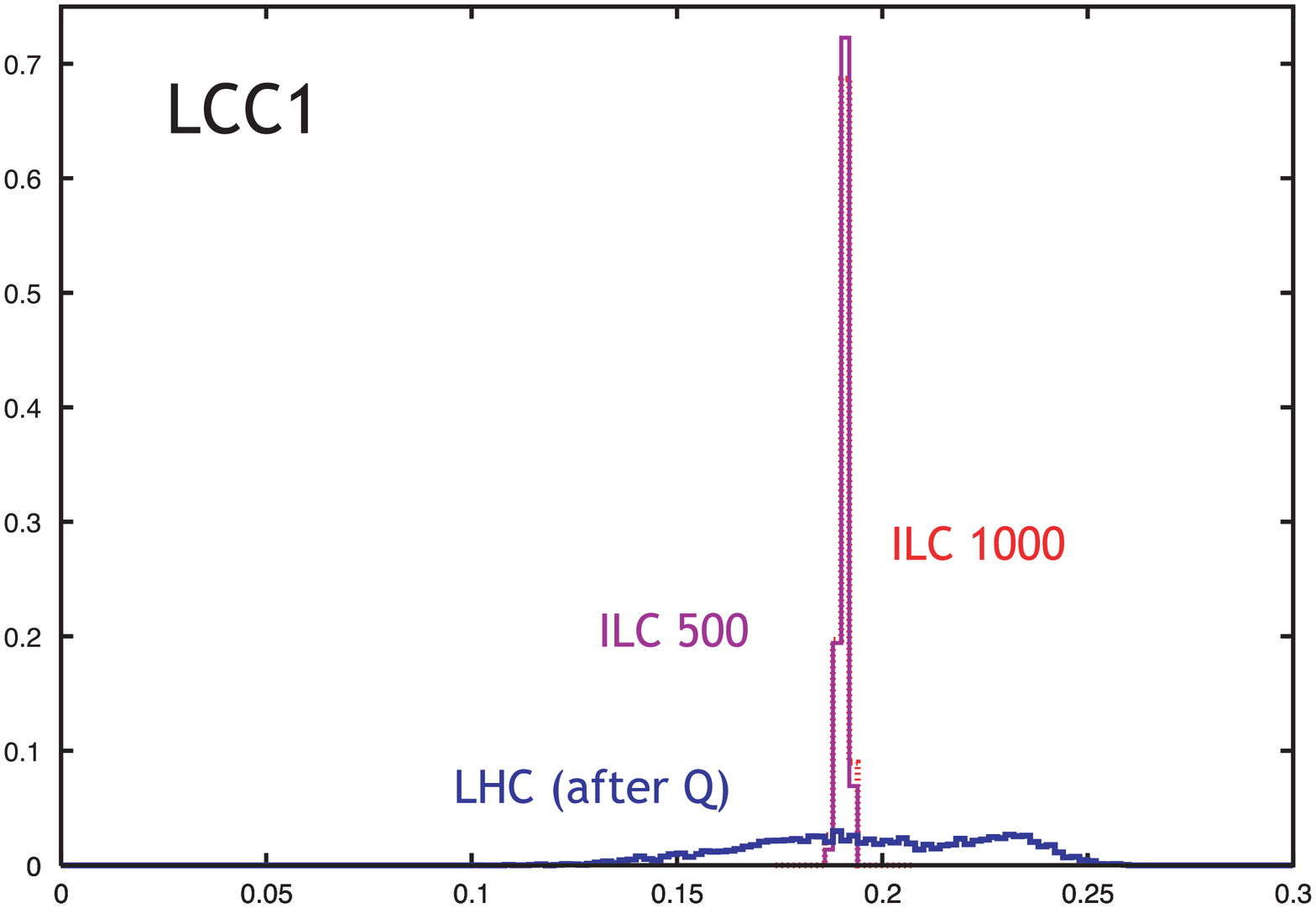}\\
 \vskip 0.2cm
\includegraphics[width=9.2cm]{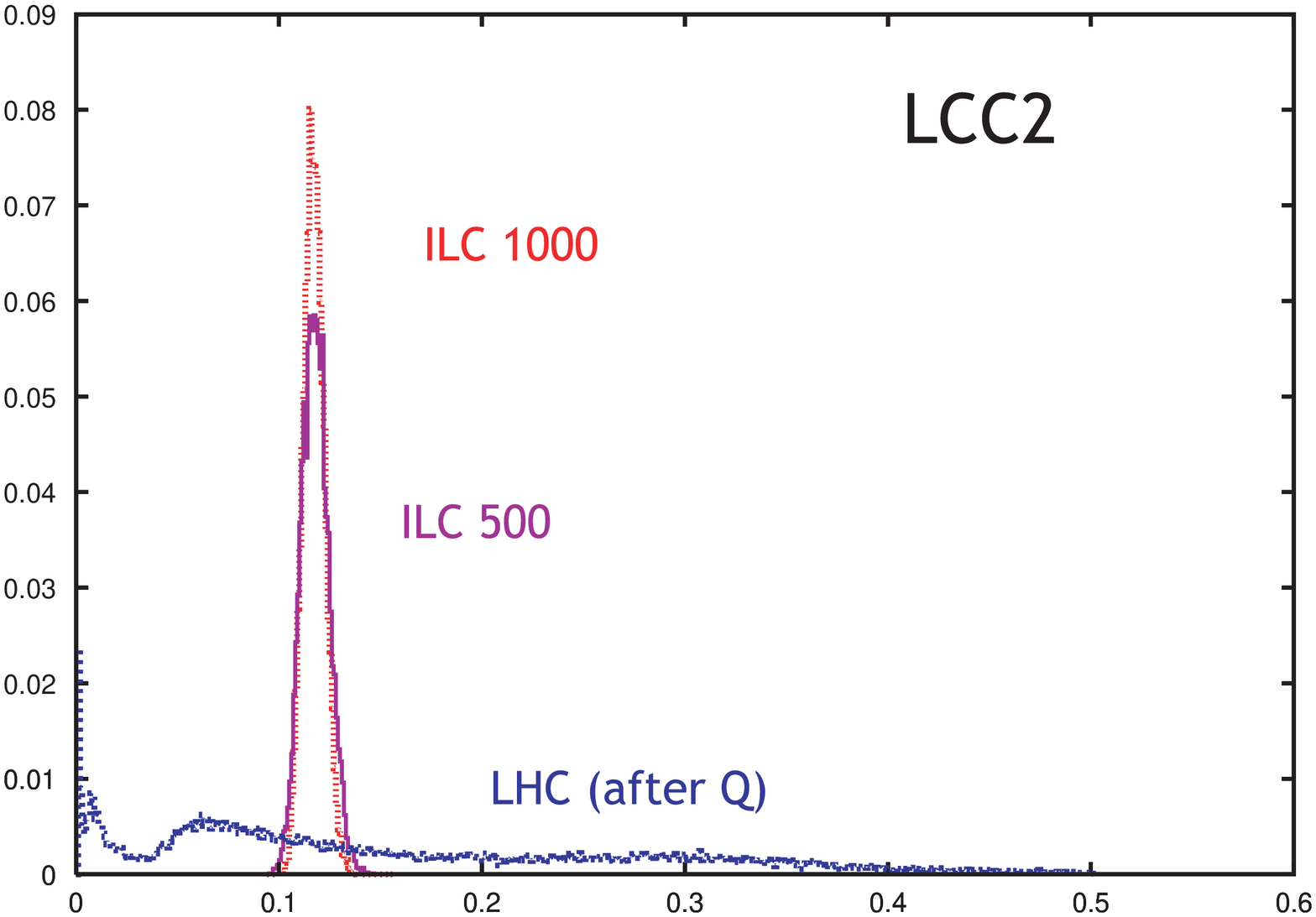}\\
\caption{Probability distribution  of predictions for
  $\Omega_\chi h^2$ from collider
measurements, using expected measurements from future colliders.
In each plot, the three distributions represent the predictions using data
 from the LHC, from the 500 GeV ILC
(peaked solid histograms), 
and the 1 TeV ILC (dotted histograms).  The 
terminology `LHC (after Q)' is explained in the text.}
 \label{fig:LCCcomp}
\end{figure*}

In Fig.~\ref{fig:LCCcomp}, we show the comparison of the determination of
$\Omega_\chi h^2$ from collider data for the reference points LCC1 and LCC2.
The figures are constructed by choosing an appropriate 
supersymmetry parameter point $P_0$, writing, for LHC and for ILC, a suite
of measurements, with errors, that would be expected in that model, and then
scanning the 24-dimensional parameter space of the flavor- and CP-conserving
MSSM to identify models consistent with this set of measurements.  Each 
model appears with a weight proportional to its likelihood, assuming 
Gaussian errors in the measurements.

The point LCC1 is identical to the point SPS1a that was studied in detail for
collider experiments in \cite{ILCLHC}.  The point is unusual in that squarks
produced at the LHC decay through a cascade of two-body decays that include
on-shell sleptons.  Identification of the endpoints gives enough kinematic
constraints to determine all of the light slepton and neutralino masses.  In
addition, an $A^0$ boson light enough to provide significant resonant 
anninhilation would be directly observed through its decay to $\tau^+\tau^-$;
thus, the presence of such a light $A^0$ can be excluded from the LHC data.
Nevertheless, the accuracy of these measurements estimates the dark matter
density only up to about 20\% accuracy.  At the ILC, the masses of light 
sleptons and neutralinos are determined to parts per mil.  This gives
much stronger constraints, and a determination of the relic density to the 
level of 1\%.
 
The point LCC2 is one in which the dominant neutralino annihilation is to 
$W^+W^-$, $Z^0Z^0$, and $Z^0 h^0$.
  In the limit that the neutralino is purely the 
supersymmetric partner of the $U(1)$ gauge boson, these annihilation reactions
are forbidden.  So the value of the annihilation cross sections is controlled
by the size of the gaugino-Higgsino mixing angles.  These must be infered
from the details of the chargino and neutralino spectrum.  In $\ee$
annihilation, we obtain additional constraints from the polarized $\ee$
pair-production cross sections.  

The point LCC2 corresponds to the  more generic situation in which the 
LHC can make a 
limited set of precision supersymmetry spectroscopy measurements.
The squarks and sleptons are very heavy at this point.
However, the gluino has a mass of about  850 GeV,
so gluino pairs are copiously produced at the LHC.  A gluino
 decays to $q\bar q$ 
plus a neutralino or chargino.  The mass difference between $\s\chi_2^0$
and $\s\chi_1^0$, and also that between $\s\chi_3^0$
and $\s\chi_1^0$, is less than $m_Z$, and so these decays contain
dilepton cascades that will allow measurement of  neutralino mass
differences to the 1\% level.
   Still, it turns out to be difficult at the LHC to exclude scenarios with 
large mixing angles or relatively light staus that lead to rapid annihilation
and low values of $\Omega_\chi h^2$. As a result, the predictions for 
$\Omega_\chi h^2$ from the LHC data cover a very broad range.
  At the ILC, measurements
of the production of the lightest charginos and neutralinos, with mass
measurements to part per mil, lead to a prediction of the relic density with
few-percent accuracy.

\begin{figure*}[t]
\centering
\includegraphics[width=11cm]{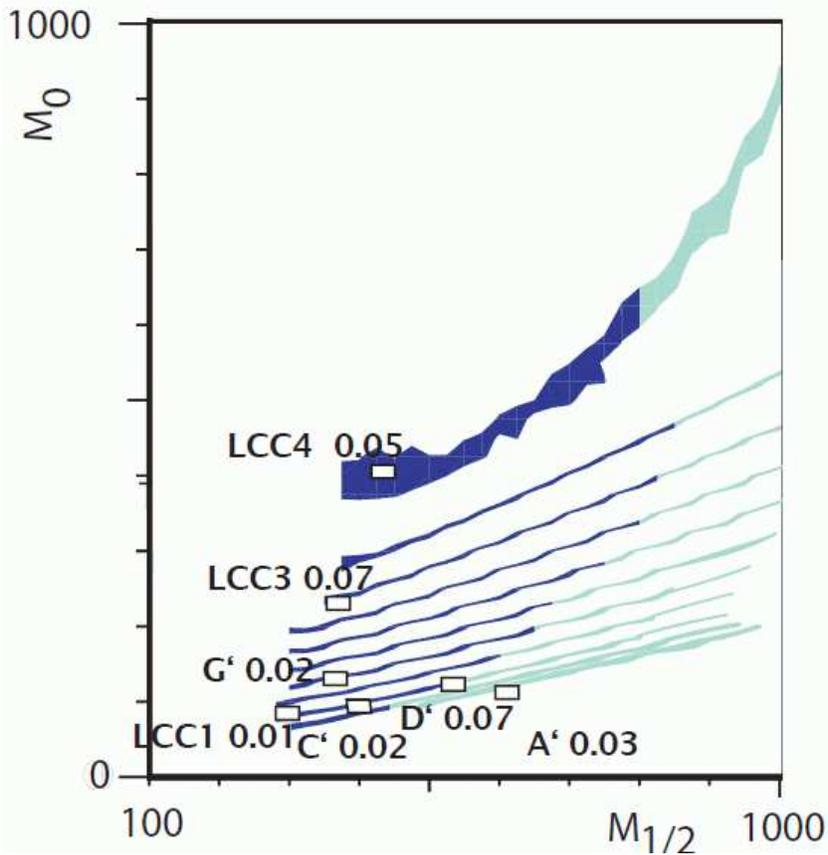} 
\caption{Estimated accuracies $\delta \Omega_\chi/\Omega_\chi$ for
 the neutralino dark matter relic density
as a function of the mSUGRA parameters.}
 \label{fig:global}
\end{figure*}

The situation is similar at other points of the MSSM parameter space.  The
determination of the WIMP annihilation cross section requires control over 
the couplings and masses of the light particles that decay to the WIMP. 
 The energy reach of the 
LHC is ultimately less important that the ILC's ability to study these 
particles in a precise way.

The variation of the error on $\Omega_\chi h^2$ from ILC
 data over the parameter
space of mSUGRA models is shown in Fig.~\ref{fig:global}. The estimates
plotted for the 
LCC points are from Table I; those for the points $A'$, $C'$, $D'$, $G'$ are 
from~\cite{Richard}.

\section{FIXING THE NEUTRALINO DETECTION CROSS SECTIONS}

If we can understand the underlying physics associated with the WIMP, we 
can also determine the cross sections for direct and indirect detection
of WIMPs in astrophysical experiments.  For these quantities, the 
comparison  of microscopic and astrophysical results brings in new issues.
Astrophysical detection rates depend on the basic cross sections, but they 
also depend on how dark matter is distributed in the galaxy.

From dynamical studies and from gravitational lensing, we now understand
that the distribution of dark matter in clusters of galaxies and on 
super-galactic scales is rather smooth, and is in accord with simulations 
of the formation of cosmic structure.  However, a lingering puzzle of 
the cold dark matter model of structure formation is that it seems to predict 
a great deal of structure in the dark matter on scales smaller than that
of the galaxy.   The cold dark matter model has been claimed to conflict
with observations in predicting a greater density of dark matter at the 
center of the galaxy than is observed, and a larger number of dwarf
galaxy companions of the Milky Way.  This situation is reviewed in
\cite{Ostr,Bertone}.  It is unclear whether 
the predictions are wrong because the simulations of dark matter in the 
galaxy are not sufficiently complete, or whether the predictions are correct
but the resulting structure is not visible to current experiments.

To obtain a first idea of the issues involved, consider the problem of 
observing dark matter in the galaxy indirectly through the flux of 
gamma rays from dark matter annihilation.  This flux is given by the formula
\beq
 E_\gamma {d \Phi\over d E_\gamma} =   E_\gamma {d (\sigma v)\over d E_\gamma}
 \cdot {1\over 4 \pi m_\chi^2} \cdot \int dz  \ \rho^2(z)
\eeq{gammarayform}
The first two factors here are essentially microscopic quantities, and we
might hope to determine them at colliders.  The last quantity, proportional
to the square of the mass density of dark matter, is determined by 
astrophysics.  Many papers pretend that this quantity can be taken as known.
But, in fact, for the dark matter density at the galactic center, this 
integral varies by five orders of magnitude, for example, among the
default density profiles of DarkSUSY~\cite{DarkSUSY}.  
It  is actually an ill-posed
problem to try to determine both the microscopic properties of dark matter and 
the distribution of dark matter from the same data set. 

The data needed to determine the detection cross sections at colliders is
similar to that needed to fix the annihilation cross section that enters the
calculation of the relic abundance.  The cross section in 
\leqn{gammarayform} is relatively straightforward to analyze:  The gamma
ray spectrum has almost the same form if WIMP pairs annihilate to 
$W^+W^-$, $Z^0Z^0$, or $q\bar q$, since in all three cases the gammas come
from $\pi^0$ decay in jets.  The main difficulty is in determining the
magnitude of the annihilation cross section.  It is tempting to put the 
value equal to that from \leqn{OmegaN}.  This is a good approximation if
the WIMPs annihilate in the $s$-wave and if co-annihilation with other states
is not important in setting the relic density.  Thus, the calculation of 
the cross section entering \leqn{gammarayform} is, to a great extent, a matter
of qualitatively distinguishing physics scenarios.  Some of these distinctions
could be made already at the LHC, and the distinctions would be
 straightforward to
recognize from the ILC data.  

For the  direct detection cross section, there is less astrophysical 
uncertainty but more uncertainty from  the microscopic physics.  Because the
cross section depends on only one power of the density, and because we 
live in a non-exceptional part of the galaxy, the uncertainty from 
astrophysics might be only a factor of two.  However, the detection
cross section can have important contributions from $s$-channel squarks, whose
properties will not be well constrained at the ILC, so it might not be 
possible to fix this cross section microscopically to the high accuracy
with which we can predict the relic density.   Still, if squarks are 
relatively light or, on the other hand, too heavy to give large 
contributions, we will be able to make firm predictions for this cross section.

It is generally the case in astrophysics that observable quantities are 
convolutions of microscopic cross sections with densities that are determined
by cosmic processes.  The study of dark matter is no different.  In
fifteen years, with the ILC data and with data from the coming generations
of underground and high-energy astrophysics experiments, we will have a large
set of varied and complementary measurements.  These may well solve the 
current questions about the distribution of dark matter in the galaxy. At the
very least, they will take us a long way from our current state of ignorance.

\section{CONCLUSIONS}

In this article, we have discussed WIMP models of dark matter and the 
possibility that we can elucidate these models experimentally.  To confront
these models with experiment, to find out whether WIMPs exist and whether they 
provide all or any of the dark matter, many steps are required.  We
must:
\begin{enumerate}
\item  Discover missing-energy events at a collider and estimate the
   mass of the WIMP.
\item  Observe dark matter particles 
   in the galaxy, and determine whether their mass
     is the same as that observed in collider experiments
\item  Determine the qualitative physics model 
          that leads to the missing-energy 
     signature.
\item  Determine the parameters of this model that predict the WIMP
             relic density.
\item Determine the parameters of this model that predict the direct and 
  indirect detection cross sections
\item   Measure products of cross sections and densities from astrophysical
 observations to build the picture of dark matter in the galaxy.
\end{enumerate}

If dark matter is composed of a single type of WIMP, this program of 
measurements should lead us to a complete understanding of what this particle
is and how it is distributed in the galaxy.  If the composition of dark 
matter is more complex, we will only learn this by carrying out this 
program and finding that it does not sum to a complete picture.  Hopefully,
further evidence from the microscopic theory will suggest other necessary
ingredients.

Both high-energy physics and astrophysics measurements are required for this 
program.  From the high-energy physics side, the first step should be
achieved at the LHC.  To make further progress, however, we will need the
capabilities of the ILC.  When the program is complete, astrophysicists will
see the ILC as a crucial tool for our understanding of the universe.

\begin{acknowledgments}
This work draws on discussions of the 
working group on ILC/Cosmology connections headed by 
Jonathan Feng and Mark Trodden, on discussions with Genevieve Belanger,
Andreas Birkedal, and Konstantin Matchev, and on results of our collaboration
with Ted Baltz and Tommer Wizansky.  We thank these
people and many others who have helped us to understand the mysteries of 
dark matter and the possibilities for their resolution.  The work of MB 
was supported by the US Department of Energy under Contract No. 
DE--AC02-05CH11231 and used resources of the National Energy Research
Scientific Computing Center, supported by Contract No. DE-AC03-76SF0098.
The work of MEP was supported by the US Department of Energy under Contract No.
DE--AC02--76SF00515. 
\end{acknowledgments}


\begin{thebibliography}{99}


\bibitem{TurnerScherrer}
 R.~J.~Scherrer and M.~S.~Turner,
  Phys.\ Rev.\ D {\bf 33}, 1585 (1986)
  [Erratum-ibid.\ D {\bf 34}, 3263 (1986)].

\bibitem{axion}
 S.~J.~Asztalos {\it et al.},
  Phys.\ Rev.\ D {\bf 69}, 011101 (2004)
  [arXiv:astro-ph/0310042].


\bibitem{Tovey}
 D.~R.~Tovey,
  Eur.\ Phys.\ J.\ direct C {\bf 4}, N4 (2002).

\bibitem{CMSCH}
  H.~C.~Cheng, K.~T.~Matchev and M.~Schmaltz,
  Phys.\ Rev.\ D {\bf 66}, 056006 (2002)
  [arXiv:hep-ph/0205314].

\bibitem{BDDKM}
  M.~Battaglia, A.~Datta, A.~De Roeck, K.~Kong and K.~T.~Matchev,
  JHEP {\bf 0507}, 033 (2005)
  [arXiv:hep-ph/0502041];
  arXiv:hep-ph/0507284.

\bibitem{SW}
  J.~M.~Smillie and B.~R.~Webber,
  arXiv:hep-ph/0507170.


\bibitem{Polesello}
 G.~Polesello and D.~R.~Tovey,
  JHEP {\bf 0405}, 071 (2004)
  [arXiv:hep-ph/0403047].

\bibitem{Battaglia:2004mp}
M.~Battaglia, I.~Hinchliffe and D.~Tovey,
J.\ Phys.\ G {\bf 30}, R217 (2004)
[arXiv:hep-ph/0406147].

\bibitem{SuperWIMPs}
  J.~L.~Feng, A.~Rajaraman and F.~Takayama,
  Phys.\ Rev.\ Lett.\  {\bf 91}, 011302 (2003)
  [arXiv:hep-ph/0302215].


\bibitem{Steigman}
 K.~A.~Olive, G.~Steigman and T.~P.~Walker,
  Phys.\ Rept.\  {\bf 333}, 389 (2000)
  [arXiv:astro-ph/9905320].

\bibitem{ILCLHC}
  G.~Weiglein {\it et al.}  [LHC/LC Study Group],
  arXiv:hep-ph/0410364.

\bibitem{Gray}
  R.~Gray {\it et al.}, in these proceedings;
  arXiv:hep-ex/0507008.
       
\bibitem{Dutta}
 V.~Khotilovich, R.~Arnowitt, B.~Dutta and T.~Kamon,
  Phys.\ Lett.\ B {\bf 618}, 182 (2005)
  [arXiv:hep-ph/0503165].

\bibitem{BattagliaParis}
  M.~Battaglia,
  arXiv:hep-ph/0410123.

\bibitem{ISAJET}
  F.~E.~Paige, S.~D.~Protopescu, H.~Baer and X.~Tata,
  arXiv:hep-ph/0312045.

\bibitem{Micromegas}
 G.~Belanger, F.~Boudjema, A.~Pukhov and A.~Semenov,
  Comput.\ Phys.\ Commun.\  {\bf 149}, 103 (2002)
  [arXiv:hep-ph/0112278];
  arXiv:hep-ph/0405253;

\bibitem{DarkSUSY}
   P.~Gondolo, J.~Edsjo, P.~Ullio, L.~Bergstrom, M.~Schelke and E.~A.~Baltz,
  JCAP {\bf 0407}, 008 (2004)
  [arXiv:astro-ph/0406204];
  New Astron.\ Rev.\  {\bf 49}, 149 (2005).

\bibitem{Seckel}
 K.~Griest and D.~Seckel,
  Phys.\ Rev.\ D {\bf 43}, 3191 (1991).

\bibitem{Richard}
 P.~Bambade, M.~Berggren, F.~Richard and Z.~Zhang,
  arXiv:hep-ph/0406010.

\bibitem{Berg}
  B.~A.~Berg,
  arXiv:cond-mat/0410490.

\bibitem{BaltzGondolo}
 E.~A.~Baltz and P.~Gondolo,
  JHEP {\bf 0410}, 052 (2004)
  [arXiv:hep-ph/0407039].

\bibitem{BBPW}
   E. A. Baltz, M. Battaglia, M. E. Peskin, and T. Wizansky, to appear.

 \bibitem{Ostr}
 J.~P.~Ostriker and P.~J.~Steinhardt,
  Science {\bf 300}, 1909 (2003)
  [arXiv:astro-ph/0306402].

\bibitem{Bertone}
  G.~Bertone, D.~Hooper and J.~Silk,
  Phys.\ Rept.\  {\bf 405}, 279 (2005)
  [arXiv:hep-ph/0404175].

\end{thebibliography}
\end{document}